\documentclass[twocolumn]{aastex631}
\definecolor{mygreen}{RGB}{0, 185, 118}

\usepackage{amsmath, amssymb, amsfonts, latexsym}        
\defcitealias{Sahade2022}{S22}
\shorttitle{Understanding the `Cartwheel CME' deflection}
\begin{document}

\title{Understanding the deflection of the `Cartwheel CME': data analysis and modeling}
\correspondingauthor{Abril Sahade}
\email{asahade@unc.edu.ar}

\author[0000-0001-5400-2800]{Abril Sahade}
\affiliation{Instituto de Astronom\'{\i}a Te\'orica y Experimental, CONICET-UNC, C\'ordoba, Argentina.} \affiliation{Facultad de Matem\'atica, Astronom\'{\i}a, F\'{\i}sica y Computaci\'on, Universidad Nacional de C\'ordoba (UNC), C\'ordoba, Argentina.}
\affiliation{Observatorio Astron\'omico de C\'ordoba, UNC, C\'ordoba, Argentina.}

\author[0000-0002-8164-5948]{Angelos Vourlidas}
\affiliation{The Johns Hopkins University Applied Physics Laboratory, Laurel MD 20723, USA.} 

\author[0000-0003-1162-5498]{Laura A. Balmaceda}
\affiliation{Heliophysics Science Division, NASA Goddard Space Flight Center, Greenbelt, MD, 20771, USA.} 
\affiliation{George Mason University, Fairfax, VA, 22030, USA.}

\author[0000-0002-9844-0033]{Mariana C\'ecere}
\affiliation{Instituto de Astronom\'{\i}a Te\'orica y Experimental, CONICET-UNC, C\'ordoba, Argentina.}
\affiliation{Observatorio Astron\'omico de C\'ordoba, UNC, C\'ordoba, Argentina.}

\begin{abstract}
    We study the low corona evolution of the `Cartwheel' coronal mass ejection (CME; 2008 April 09) by reconstructing its 3D path and modeling it with magneto-hydrodynamic simulations. This event exhibited a double-deflection that has been reported and analyzed in previous works but whose underlying cause remained unclear. The `Cartwheel CME' traveled toward a coronal hole (CH) and against the magnetic gradients. Using a high-cadence, full trajectory reconstruction, we accurately determine the location of the magnetic flux rope (MFR) and, consequently, the magnetic environment in which it is immersed. We find a pseudostreamer (PS) structure whose null point may be responsible for the complex evolution of the MFR at the initial phase. From the pre-eruptive magnetic field reconstruction, we estimate the dynamic forces acting on the MFR and provide a new physical insight on the motion exhibited by the 2008 April 09 event. By setting up a similar magnetic configuration in a 2.5D numerical simulation we are able to reproduce the observed behavior, confirming the importance of the PS null point. We find that the magnetic forces directed toward the null point cause the first deflection, directing the MFR towards the CH. Later, the magnetic pressure gradient of the CH produces the reversal motion of the MFR.  
\end{abstract}

\section{Introduction} 




Coronal mass  ejections (CMEs) are the drivers of the strongest geomagnetic storms and a major concern of space weather. They are usually related to the ejection of magnetic flux rope a (MFR) that connects them to the eruptive source region in the lower corona, including prominence/filament eruptions, flares, and cavities \citep[e.g.,][]{Zhang2001,vanDriel2015,Green2018,Jiang2018,Yang2018,Filippov2019}.  Predicting the  occurrence and trajectory of the eruption is crucial for assessing their potential geoeffectiveness. Since the launch of the \textit{Solar TErrestrial RElations Observatory} \citep[STEREO,][]{kaiser2008} twin spacecrafts (STA and STB, hereafter), along with the development of various reconstruction tools \citep[e.g.,][]{mierla2008, maloney2009, temmer2009, Thernisien2009,Kwon2014,Isavnin2016,Zhang2021A&A}, multi-point observations allow the determination of the three-dimensional (3D) path of CMEs and their associated source regions.  

Several factors can deflect an eruption from its radial course \citep{macqueen1986,cremades2004,gui2011,kay2015,Sieyra2020}. It is generally accepted that neighboring magnetic structures, such as coronal holes \citep[CHs - e.g.,][]{Cremades2006,Gopalswamy2009,Sahade2020,Sahade2021} and active regions \citep[ARs - e.g.,][]{kay2015,Mostl2015,Wang2015}, can deflect MFRs in longitude and latitude against their position. On the other hand, heliospheric current sheets \citep[e.g.,][]{Liewer2015,Wang2020JGRA}, helmet streamers \citep[e.g.,][]{Zuccarello2012,Yang2018}, and pseudostreamers \citep[PSs - e.g.,][]{cecere2020,Wang2020JGRA, Karna2021, Sahade2022} attract MFRs toward their low magnetic field regions. This responses can be quantified, in strength and direction, by the local and global gradients of the magnetic pressure \citep{gui2011,Panasenco2013SoPh,Liewer2015, Sieyra2020}. 

However, there are events that seem to propagate against those gradients, such as the one known as `Cartwheel CME.' This event erupted on 2008 April 09, after 8:45 UT, and has been studied extensively from different perspectives
\citep{Landi2010ApJ,Savage2010,gui2011,Patsourakos2011,Thompson2012,Kliem2012,Capannolo2017}. The eruption followed a non-radial trajectory, according to 3D reconstructions \citep{Landi2010ApJ,gui2011,Patsourakos2011,Thompson2012}. \citet{Landi2010ApJ} first reconstructed the 3D CME core trajectory at eight different times from $1.1$ to $5.1\,R_\sun$ noting that the `Cartwheel CME' had an initial deviation toward the Earth and later it moved away from Earth direction. \citet{Savage2010} tracked the erupted material in STA plane-of-sky (POS) with better cadence but in a 2D projection below $ 1.5\,R_\sun$. They investigated the magnetic field configuration noting that the CME seems to initially move toward the southern open field lines. Later, the trajectory projected in the POS becomes more radial near $\sim 2.5\,R_\sun$. To understand the non-radial evolution of this event, \citet{Capannolo2017} modeled the MFR eruption with ForeCAT \citep{Kay2013,kay2015} and compared it to the reconstructed trajectory of \citet{Landi2010ApJ}. ForeCAT calculates the deflection and rotation of the simulated MFR (varying initial mass, speed, size, shape, and location) considering the magnetic forces (tension and pressure gradient) from the solar background. Although they were able to reproduce the double deflection, they found that the MFR moved unexpectedly against the magnetic gradients, toward a CH. They needed to assume a non-radial initial velocity to impulse the MFR in this direction and proposed that an asymmetrical reconnection of the footpoints could explain it.

In this paper, we investigate the validity of the previous interpretations about the deflection of the `Cartwheel CME' and find that the eruption is not unusual but follows the expected trajectory along existing magnetic fields. A detailed reconstruction allows to properly investigate the magnetic interaction between the environment and the `Cartwheel CME' and provide a new insight into the MFR behavior. In Section~\ref{sec:data}, we reconstruct the 3D path of the 2008 April 09 event with higher cadence and using different techniques at the low corona. We reconstruct the surrounding magnetic field with the Potential Field Source Surface model \citep[PFSS,][]{2003SoPh..212..165S}. In Section~\ref{sec:simu}, we present the results of a magnetohydrodynamic (MHD) numerical simulation where a MFR interacts with the main magnetic structure found by the PFSS reconstruction. The simulated event reproduce the `Cartwheel CME' behavior and allow us to compute the forces acting on the MFR. Conclusions and final comments are presented in Section~\ref{sec:conclusion}.

\section{Data analysis} \label{sec:data}
\subsection{Source region}
\begin{figure}    
   \centerline{\includegraphics[width=0.45\textwidth,clip=]{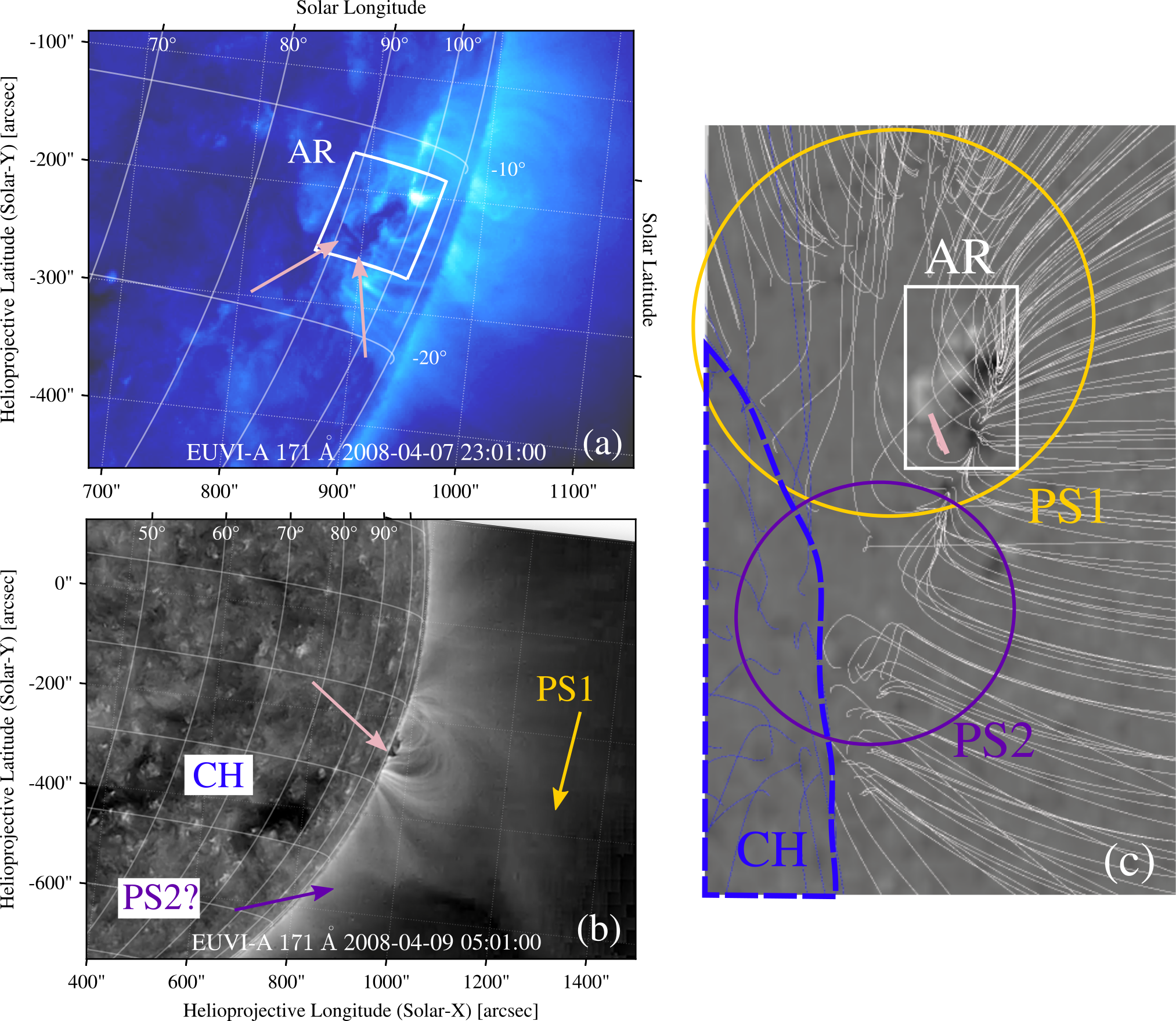}
             }
 \caption{(a) Source region and pre-eruptive filament (pink arrows) seen by STA on 2008 April 07. The white square indicates the AR as in the right panel. (b) Pre-eruptive filament (pink arrow) seen by STA on 2008 April 09 hours before the eruption, PS1 is located behind limb (orange arrow), the CH is seen as a dark patch, and some field lines of PS2 (violet arrow) are possibly seen at the limb. (c) Radial magnetic field $B_r$ at $1 R_\sun$ in gray-scale, open (blue) and closed (white) field lines, position of the pre-eruptive filament (pink line), and magnetic structures surrounding the source region. The white square indicates the AR 10989,the pseudostreamers are indicated by circles (in orange PS1, in violet PS2) and the southern CH is delimited by blue dashed lines.} 
\label{fig:sourceregion}
\end{figure}

Much of the eruptive material belonged to a prominence located within AR 10989. Figure~\ref{fig:sourceregion} shows the source region in 171 \AA from STA perspective and the magnetic structures near it. Figure~\ref{fig:sourceregion}(a) shows the pre-eruptive filaments on 2008 April 07, when the AR was still on disk. The eruptive filament was not the one lying along the polarity inversion line (PIL) in the center of the AR, but the one in the outer part of it (pink arrows). Figure~\ref{fig:sourceregion}(b) shows also the pre-eruptive filament (pink arrow), the position of a pseudostreamer that overlaid the AR (PS1, orange arrow), the position of the southern CH (blue label), and indicates some lines visible on the limb that may belong to a second pseudostreamer (PS2, violet arrow), all on 2008 April 09 few hours before the eruption. Figure~\ref{fig:sourceregion}(c) shows the radial magnetic field $B_r$ at $1 R_\sun$ in gray-scale, the open (blue) and closed (white) field lines over it, the position of the pre-eruptive filament (pink line) of panel (a), and the mentioned magnetic structures surrounding the source region. The white square indicates the AR 10989, the pseudostreamers are indicated by circles (in orange PS1, in violet PS2) and the southern CH is delimited by blue dashed lines. The prominence was enclosed by a PS with anemone-like topology \citep[e.g.,][PS1, see Figure~\ref{fig:sourceregion}]{Mason2022}, whose southern side was overlaid by the negative open field of a CH (see Figure~\ref{fig:sourceregion}[c]) and the northern side was overlaid by the negative footpoints of closed field lines. The region was complex and presented more PS structures, such as PS2 (see Figure~\ref{fig:sourceregion}[c]).  The PS topology consists of a separatrix dome above a minority polarity region, and an outer spine emanating from a null point on this dome and connected out into the open heliosphere or to some far distant closed-field region. This is the well-known embedded-bipole topology surrounded by unipolar fluxes of both open or closed (at larger scales) magnetic fields \citep[e.g.,][]{Raouafi2016,Mason2021,Wyper2021}.

Between 2008 March 22 and 30, the AR exhibited eruptive activity, with a major CME on 2008 March 25. After that, the region remained quiet until 2008 April 03, when it exhibited brightening in the EUV 195 channel and two small eruptions on 2008 April 05. The `Cartwheel CME' is the last and most notable of the eruptions from this region.  

\subsection{Prominence and CME 3D Reconstruction}
\begin{figure}    
   \centerline{\includegraphics[width=0.45\textwidth,clip=]{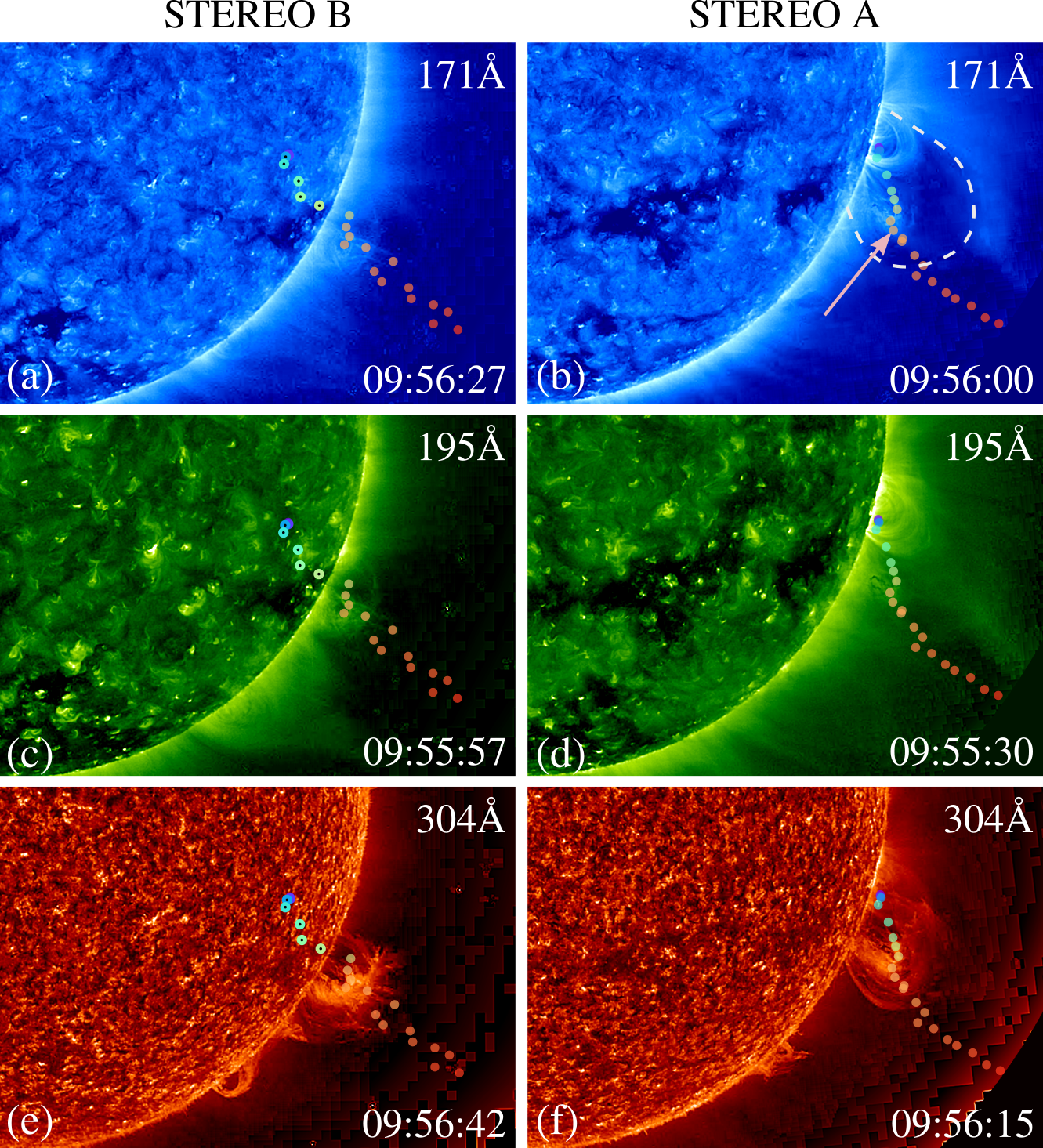}
             }
 \caption{2008 April 09 event in STEREO 171, 195 and 304 \AA filters at 09:56 UT. Rainbow-color dots represent the apex position of the prominence from 8:25 UT (violet dot) to 10:40 UT (red dot), triangulated in EUVI images. Left panels show STB FOV, dots with black center are behind limb and are triangulated with SOHO/EIT; right panels show STA FOV. Panel (b) indicates also the prominence position with the pink arrow, and the CME front in white dashed lines.}
\label{fig:EUVI-triang}
\end{figure}

The 2008 April 09 event was observed on the west limb by the \textit{Solar and Heliospheric Observatory} \citep[SOHO,][]{1995SoPh..162....1D} and STEREO spacecraft. At that time, the STEREO spacecraft were separated by $\sim24^{\circ}$ from Earth. We use the data provided by the \textit{ Extreme ultraviolet Imaging Telescope} \citep[SOHO/EIT,][]{1995SoPhEIT}, the \textit{Large Angle and Spectrometric Coronagraph Experiment} \citep[SOHO/LASCO,][]{1995SoPhLASCO}, the wavelet-enhanced images from \textit{Extreme-Ultraviolet Imager }\citep[STEREO/EUVI,][]{Howard2008}, and COR1 coronagraphs from STEREO spacecraft to reconstruct the trajectory of the prominence and CME. We use \textit{Michelson Doppler Imager} \citep[SOHO/MDI,][]{1995SoPh..162..129S} data for the days before 2008 April 09 and apply the PFSS model to reconstruct the magnetic field over the solar surface.

Since the source region was located near the western limb of STEREO-A (STA), we reconstruct the initial 3D trajectory from SOHO/EIT 195 and STA/EUVI 195 \AA channels. When the prominence appears in the STEREO-B  field-of-view (FOV), we track the ejected material in the 171, 195 and 304 \AA channels, from both STEREO spacecraft to ensure we are following the same features and cover the broader time range with high cadence. Finally, we track the prominence in white-light images from STEREO/COR1 while it is bright and compact. The 3D location of the prominence is determined using the tie-pointing technique, which consists of a geometrical reconstruction by considering the position of the same feature in the FOV of two different spacecraft \citep[see, e.g.,][]{Inhester2006}. We use the \texttt{scc$\_$meassure} routine, developed by B. T. Thompson, from \textit{SolarSoft}. Figure~\ref{fig:EUVI-triang} shows the eruption at 09:56 UT from STA and STB perspective in the different filters. For the 171 and 195 \AA filters we follow the apex of the cold material prominence. In the 304 \AA filter and COR1 images, we track the main axis of the prominence, measuring multiple positions each time.  The median latitudes and longitudes correspond to the position of the apex.  The color-coded dots in Figure~\ref{fig:EUVI-triang} summarize the reconstructed trajectory of the prominence apex at each time. Figure~\ref{fig:EUVI-triang}(b) also indicates the prominence position (pink arrow) and CME front (white dashed lines) at that time.

\begin{figure}    
   \centerline{\includegraphics[width=0.45\textwidth,clip=]{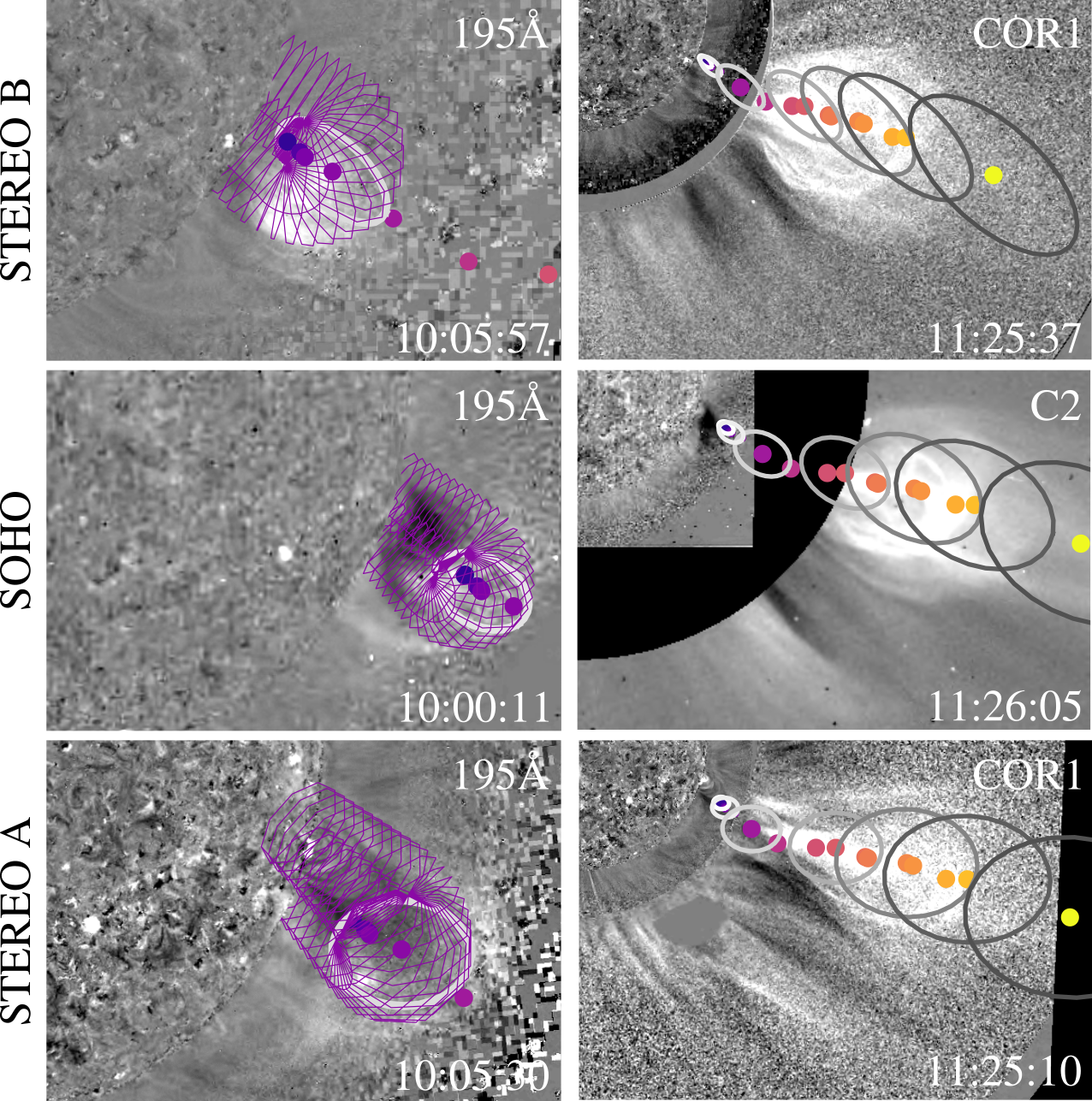}
              }
 \caption{2008 April 09 event in STA, SOHO and STB in 195 \AA filter (left) and coronagraph (right) images. Images are processed with base difference. Color dots represent the position of the CME center from 9:25 UT (violet dot) to 11:45 UT (yellow dot). The GCS reconstruction at 10:05~UT (magenta wire-frame) and its central cross section (white circle) is plotted over the 195 images. The cross sections (grey circles) of the GCS reconstruction and centers (color dots) in 9:25 - 11:45~UT are over-plotted on the coronagraph composites in the right panels. 
}\label{fig:GCS}
\end{figure}

In addition, we reconstruct the CME from the EUV 195 \AA and white-light images from the three viewpoints (SOHO, STA and STB). To reproduce the evolution of the deflecting CME from the low corona we use, for the first time, a non-radial Graduated Cylindrical Shell (GCS) model. In this way, the CME footpoints coordinates can be fixed while the CME front can vary in latitude and longitude. We use the \textit{SolarSoft} routine \texttt{rtcloudwidget}, the parameters tuned for the reconstruction are: latitude, longitude, tilt angle, height, half angle, ratio and non-radial tilt. The latter allows to change the angle subtended by the CME axis and the radial plane defined by the footpoints and the solar center. This adds a degree of freedom in the reconstruction, which may produce a new set of solutions. However, as the footpoints are characterized by the latitude, longitude, and tilt angle these parameters remain unchanged throughout the full evolution of the CME. This is an improvement over previous reconstructions as the deflection can be better captured and the coordinates of the CME front are more accurately determined while CME footpoints remain in the source region. Figure~\ref{fig:GCS} shows the projected centers of the CME modeled by the GCS in color dots (from 9:25 UT in violet to 11:45 UT in yellow), left panels show the base difference images of 195 \AA filters (for STA, SOHO, and STB) and the GCS wire and cross section circle for 10:05 UT, right panels show the base difference of white-light images (also for STA, SOHO, and STB) including the projected cross section circles of the GCS model. Table~\ref{tbl-gcs} shows the parameters used for the non-radial GCS reconstruction and  Table~\ref{tbl-coord} (Appendix~\ref{app:tabla}) shows the coordinates for both GCS and triangulation reconstruction techniques, for the prominence and the CME, in EUV, and white-light images, respectively.  The prominence is tracked from 08:15 UT to 11:15 UT ($1.03-3.08\,R_\sun$), and the CME can be modeled from 09:25 UT to 11:45 UT ($1.24-4.1\,R_\sun$).

\begin{deluxetable}{c | c c c}
\tablecaption{Non-radial GCS parameters.\label{tbl-gcs}}

\tablehead{
\colhead{Time}  & Height & Aspect ratio & Non-radial tilt\\
\colhead{[UT]}& \colhead{[$R_\sun$]}& \colhead{ }& \colhead{[$^\circ$]}
}

\startdata
  09:25& $1.34$& $0.05$& $-33.5$ \\
  09:35& $1.35$& $0.06$& $-34.2$ \\
  09:45& $1.42$& $0.08$& $-34.2$ \\
  09:48& $1.43$& $0.08$& $-34.2$ \\
  10:00& $1.46$& $0.09$& $-34.8$  \\
  10:05& $1.56$& $0.12$& $-29.8$  \\
  10:14& $1.84$& $0.14$& $-29.8$  \\
  10:26& $2.11$& $0.14$& $-25.5$  \\
  10:40& $2.38$& $0.14$& $-15.5$ \\
  \hline
  10:45& $2.57$& $0.16$& $-11.2$  \\
  10:55& $2.92$& $0.19$& $-9.3$  \\
  11:05& $3.30$& $0.20$& $-5.6$ \\
  11:10& $3.37$& $0.20$& $-5.6$ \\
  11:20& $3.70$& $0.20$& $-5.6$ \\
  11:25& $3.88$& $0.20$& $-3.7$ \\
  11:45& $4.91$& $0.20$& $-3.7$  
\enddata
\tablecomments{Height parameter corresponds to the CME front height, the aspect ratio indicates the relation between height and the cross section of the CME, and the non-radial tilt gives the angle between the radial direction and the CME axis. The other parameters of the model remain fixed: latitude $\theta=-17^\circ$, longitude $\phi=196^\circ$, tilt $\gamma=-42.5^\circ$, and half angle $\alpha=4.5^\circ$. Times up to 10:40 UT correspond to EUV images, times after 10:45 correspond to coronagraph images.}
\end{deluxetable}

\begin{figure*}    
   \centerline{\includegraphics[width=0.95\textwidth,clip=]{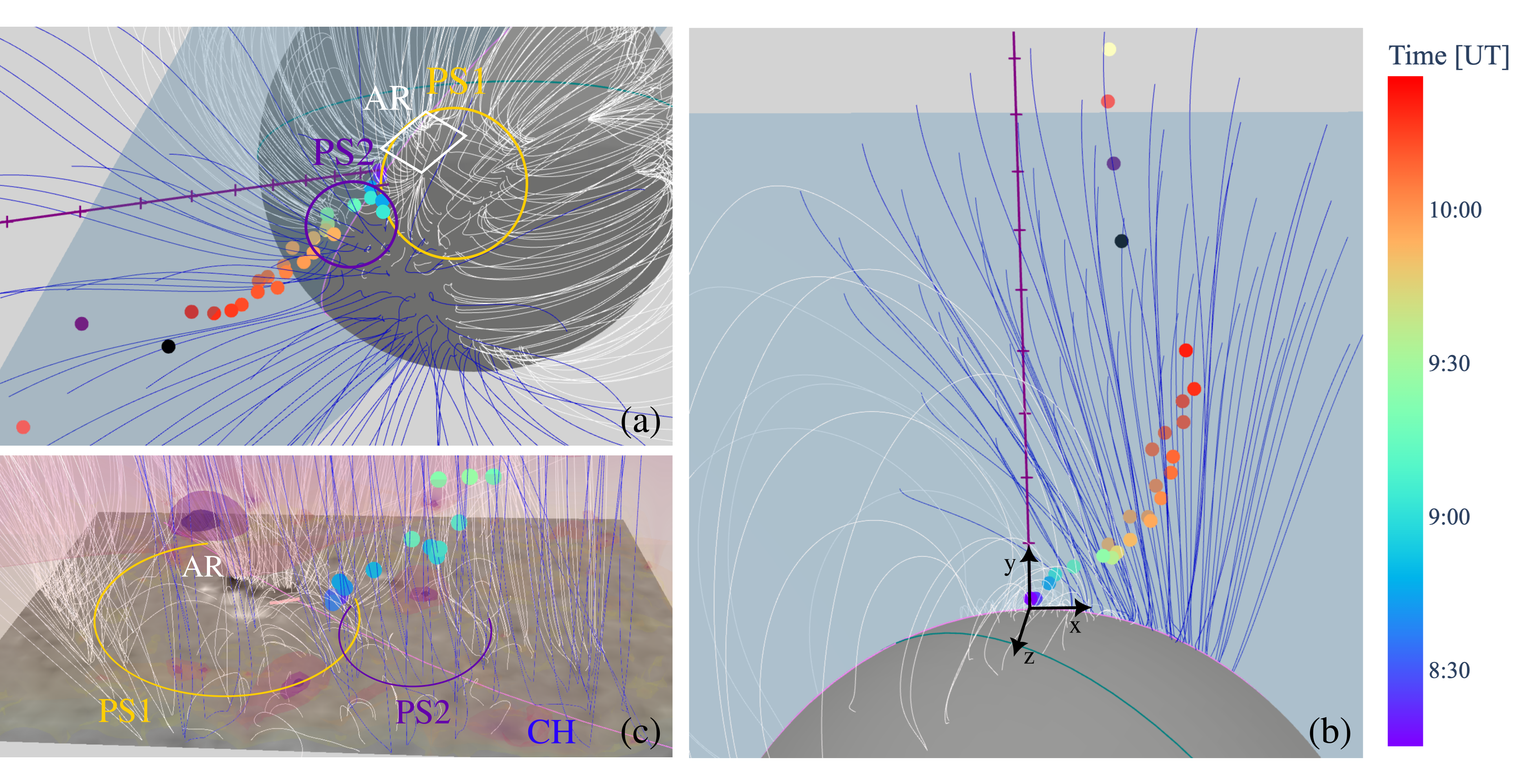}
              }
 \caption{(a) Triangulated trajectory of the prominence apex with the PFSS magnetic field lines. White magnetic field lines are closed and blue ones are open lines of positive polarity. Rainbow-color dots show the prominence triangulation in EUVI images. Black, violet, pink and beige dots are triangulated in COR1 images at 10:45 UT, 10:55 UT, 11:05 UT and 11:15 UT, respectively. In purple, the radial direction according to initial position of the prominence, with markers separated by $0.2\,R_\sun$. In light blue, the plane where the trajectory lies, the pink circle shows the intersection of this POE and solar surface, the teal circle represents the equator. Magnetic structures PS1, PS2, AR and CH idem Figure~\ref{fig:sourceregion}. (b) Rotated position of the eruption and magnetic field lines. Cartesian axes defined from the POE.(c) Radial magnetic field $B_r$ at $1 R_\sun$ in gray-scale with iso-contours of magnetic field strength in purple shades. The spots of stronger purple show the 3D location of minimum magnetic field. The magnetic fields lines are plotted in blue (open) and white (closed), also the position of the pre-eruptive filament (light-pink line), and the pink circle showing the POE projection over the surface  of the plane. The magnetic structures are the same as in panel (a).}
\label{fig:3d-view}
\end{figure*}
Figure~\ref{fig:3d-view} shows the triangulated prominence trajectory and the surrounding PFSS magnetic field lines. Again, the main magnetic structures near the source region are indicated (PS1, orange circle; PS2, violet circle, AR, white square, CH, blue lines).  The prominence initially departed from a complex region of closed loops (white lines) at Carrington longitude and latitude ($195^\circ$,$-16^\circ$). It traveled towards the open magnetic field of a southern CH (blue lines) until reaching ($182^\circ$,$-29^\circ$). Afterwards, the prominence changed its motion and traveled outward along the open magnetic field lines, with coordinates ($196^\circ$,$-26^\circ$) in the last measured position. The purple line marks the radial direction from the initial position of the eruptive material, with ticks from  $1.2$ to  $2.8\,R_\sun$. The prominence apex changed in both latitude and longitude, but it is possible to define a plane intersecting the solar sphere which contains the evolution of the apex (hereafter, plane of eruption - POE). The POE is selected by non-linear least-square fitting and presents a standard deviation lower than $1^\circ$, indicating that the apex moved within a plane (light blue plane of Figure~\ref{fig:3d-view}). Figure~\ref{fig:3d-view}(b) shows a rotated view of the eruption in which the POE is parallel to the POS and the radial direction is pointing upwards. We define a Cartesian reference system with the $x$-axis being parallel to the solar surface at the initial position of the prominence, the $y$-axis pointing in the radial direction, and $z$-axis perpendicular to the POE. In this system, the outward motion is projected in the $y$-direction, and the deflections in $x$-direction. For reference, the solar equator is shown in teal color. Figure~\ref{fig:3d-view}(c) shows a closer view of the source region and initial triangulated position of the prominence. It includes the radial magnetic field at $1\,R_\sun$ and the iso-contours of magnetic field strength to show the 3D location of null points. Null points are located in the purpule contours. In particular, we notice the one corresponding to PS1 (orange circle) and PS2 (violet circle). From this view we see how the projection of the POE (pink line) crosses the different structures, such as the external border of the AR, a lobe of PS1, PS2 and the CH. In light-pink is drawn the pre-eruptive filament.
\begin{figure}    
   \centerline{\includegraphics[width=0.49\textwidth,clip=]{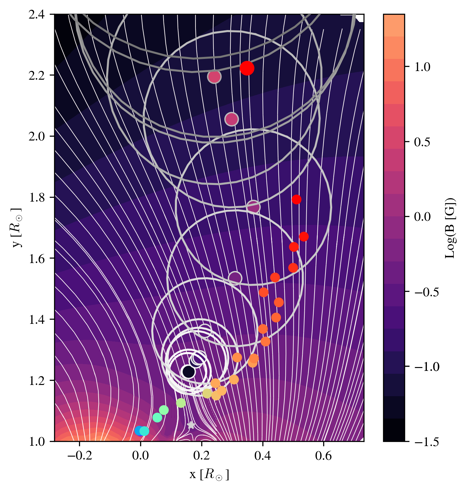}
              }
 \caption{Magnetic field magnitude (logarithmic scale) and field lines with the prominence and CME trajectory in the POE. Gray star indicates the null point position of PS2. Rainbow-color dots correspond to the triangulated positions of the prominence with same time scale as Figure~\ref{fig:3d-view} and the last larger red dot being the black one use on Figure~\ref{fig:3d-view}. White to gray circles are the cross sections of the non-radial GCS model from 9:25 UT to 11:20 UT, and color dots with same white/gray edges are the centers of each cross section, up to 10:45 UT.}\label{fig:B-triang}
\end{figure}

By defining the POE we can study the magnetic scenario that produces the non-radial motion in a simpler way as we reduce the dimension of the problem. We assume that the forces along the third dimension of this projection ($z$-axis) are balanced because the system did not suffer displacements in that direction. Figure~\ref{fig:B-triang} shows the magnetic field magnitude in logarithmic scale, the magnetic field lines and the prominence position (rainbow dots, obtained by triangulation), the CME center position (magma dots with gray edges, obtained from the non-radial GCS model) and the CME cross section (gray circles, obtained from the non-radial GCS model) projected on the POE, in the Cartesian reference system described above. The PS located at $x\sim 0.2\, R_\sun$ is PS2. To the left of PS2 there are the closed field lines belonging to the helmet streamer that encloses PS1, the orange magnetic field intensity at $x\sim -0.2\, R_\sun$ belongs to the outer part of the AR. To the right of PS2 there are the open field lines of the CH. From the early reconstruction of the prominence path, we can see that it headed toward the null point (gray star) of PS2 located at $(x,y)=(0.16\,R_\sun,1.05\,R_\sun)$, then both the CME and the prominence moved to the right (in this coordinate system) displacing $\sim 15^\circ$ from the radial direction. About $1.8\,R_\sun$ they reversed the motion traveling to the left and aligning with the CH field lines. The final angle of deflection is lower than $5^\circ$. This double-deflection behavior was previously reported in \citet{Sahade2021}. Their scenario did not include a PS configuration as in the case here, but the interaction with the PS null point and the open magnetic field lines is quite similar (see \S~\ref{sec:simu} for further analysis). It is interesting to note the evolution of the prominence relative to the CME. Initially, the prominence was located close to the right edge of the MFR cross section, exhibiting a larger deflection than the MFR center (the maximum being $18^\circ$ and $14^\circ$, respectively). In the later stages, the prominence apex progressively reached the MFR center, in both displacement $x$ and height $y$. This behavior appears consistent with the prominence material lying at the bottom of the MFR due to the gravity and the balance of magnetic forces \citep[e.g][]{Vourlidas_etal_2013}. As the MFR moved non-radially, the prominence followed along the edge of the cavity, experiencing larger deflection possibly because of its larger inertia. The displacement between the MFR and the prominence is noted in the STA images (see Figure~\ref{fig:EUVI-triang}[b], where the position of the prominence it was not centered with the CME front) but 3D measurements give us certainty that the actual trajectories differ and that it is not a projection effect. It may be due to changes in the MFR shape (i.e. contraction) as the prominence material expanded and/or drained out. It is difficult to reach robust conclusions without detailed information on the prominence physical properties as it evolves which is unavailable. 

\subsection{Magnetic forces}

\begin{figure}    
   \centerline{\includegraphics[width=0.4\textwidth,clip=]{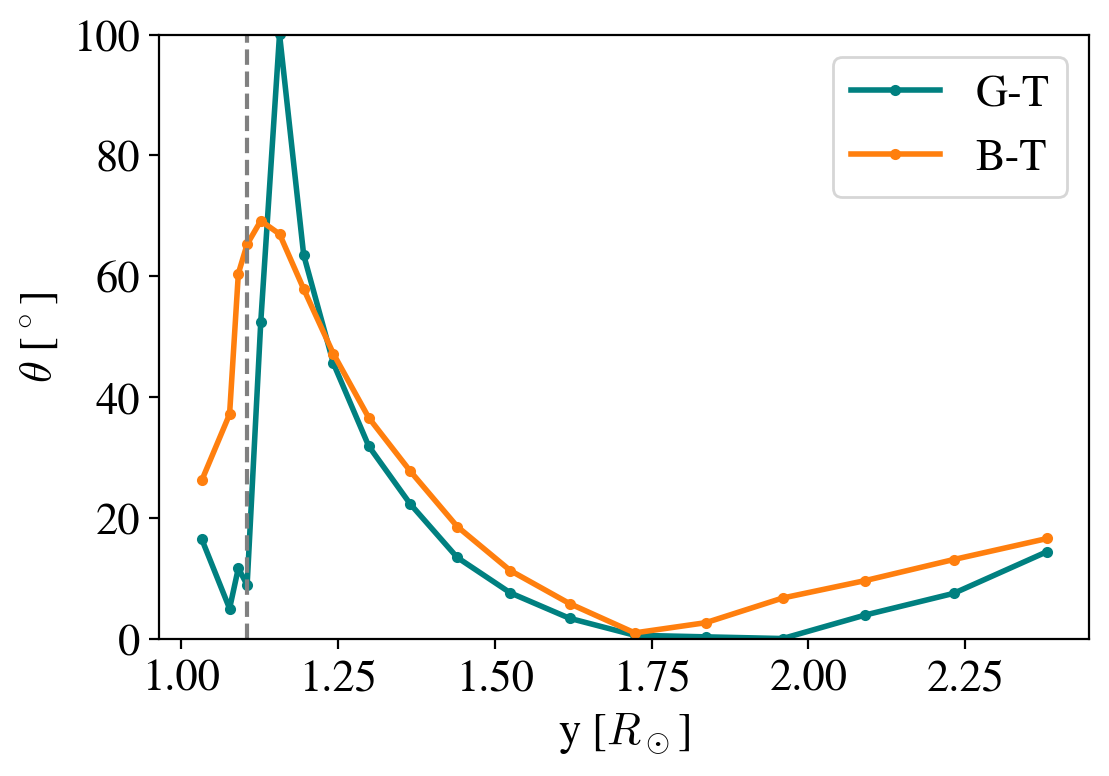}
              }
 \caption{Angle of miss-alignment between the magnetic gradient and trajectory (G-T), and the magnetic field and trajectory (B-T). The grey dashed line indicates the height at which the PS spine is crossed.}\label{fig:angle}
\end{figure}
The PFSS model is useful for understanding the global magnetic environment and large-scale structures surrounding the eruptive material. However, it cannot account for the magnetic field evolution during an eruption unless the eruption produces photospheric changes, which is observed only in large eruptions. From this reconstruction technique, we recover a PS null point which may be ``attracting'' the MFR and directing it toward the open magnetic field lines of the nearby CH. Figure~\ref{fig:angle} shows the temporal evolution of the angular alignment between the MFR trajectory and both the magnetic field lines (B-T orange line) and the gradient of magnetic pressure (with the conventional minus sign in front, i.e., $\vec{G}=-\nabla \frac{B^2}{2\mu_0}$; G-T teal line). In the initial phase of the eruption (8:45-9:35 UT; below $1.2\,R_\sun$), the MFR moved slightly misaligned with the gradient direction, but since the MFR did not stop in the PS2 null point the miss-alignment grows to $100^\circ$. In the second phase (until 10:35 UT and $1.8\,R_\sun$), the MFR moved along the CH magnetic field lines, aligning with both the gradient and field as it lost speed in the $x$-direction. In the third phase of the evolution, the misalignment remains small but with an increasing trend. This can be understood as the dynamical response of the CH, which was compressed by the inertial motion of the MFR and later returned the MFR to the original position of the open field lines. At 11:00 UT, and above $2.5\,R_\sun$, the CME stopped the $x$ displacement, confined in the lines of the CH.

\begin{figure}    
   \centerline{\includegraphics[width=0.4\textwidth,clip=]{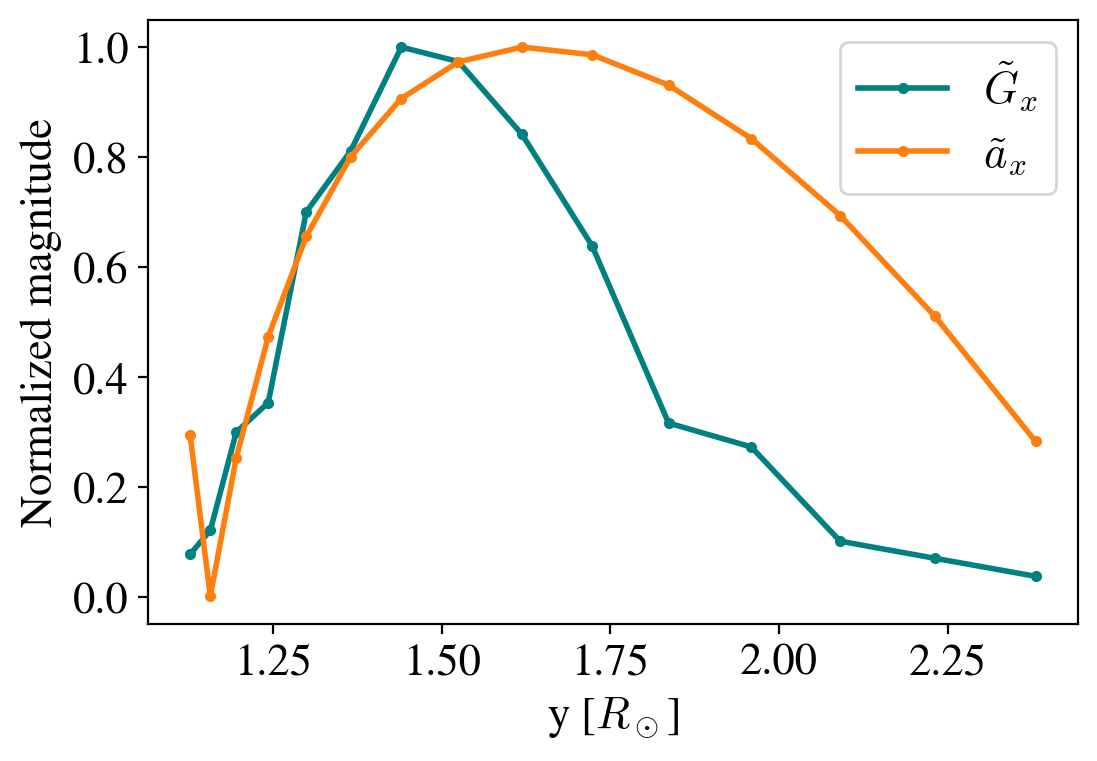}
              }
 \caption{Normalized magnetic gradient pressure ($\tilde{G}_x$) and normalized acceleration ($\tilde{a}_x$) in $x$-direction during the observed interaction with between the MFR and the CH.}\label{fig:ax-gradp}
\end{figure}

To estimate the force exerted by the CH on the MFR, we consider flux conservation of the CH magnetic field lines. Without magnetic reconnection, the field lines should be pushed inwards, reducing the CH area and proportionally increasing it magnetic field strength in the $B_y$ component. Considering this, we estimate the magnetic pressure gradient in the $x$-direction. We also obtain a polynomial fitting for the trajectory, and derive the radial ($y$-direction) and non-radial ($x$-direction) velocity and acceleration. Figure~\ref{fig:ax-gradp} compares the normalized magnitude of the force produced by the magnetic pressure gradient and the normalized magnitude of the MFR acceleration in the $x$-direction. In this evolution period the acceleration is directed to the $-x$-direction and increases until $\sim 1.7\,R_\sun$ (10:20 UT), then gradually reduces to zero. Since the trend of both curves is quite similar we suspect that the magnetic pressure gradient is contributing to accelerate the MFR out of the CH during their interaction.

\section{Numerical simulation}\label{sec:simu}

 In \citet[hereafter S22]{Sahade2022} we modeled a MFR immersed in a PS magnetic field and studied the dynamical interaction between both structures while changing the parameters. For this work, we adapt the model used in that work to simulate the magnetic configuration of the 2008 April 09 event. The `Cartwheel CME' seems to interact with the null point of a nearby small PS (PS2) and then with the CH overlying the southern lobe of that PS. By adjusting the model parameters and modifying equations (11a-b) of \citetalias{Sahade2022} to obtain a bent PS spine, we establish an initial magnetic field configuration that has a topology and magnetic field strength similar to those shown in Figure~\ref{fig:B-triang}. The new equations for the background magnetic field allow a shift of the central position of the potential magnetic field overlying the PS:

\begin{align}\label{e:BfieldPSx}
    B_{x}(x,y)=&\frac{2 \sigma  B_\text{PS} (x - x_\text{PS}) (y - y_\text{PS})}{((x - x_\text{PS})^2 + (y - y_\text{PS})^2)^2} + \\ \nonumber
    & B_0\ \sin\left(\frac{x-x_\text{CH}}{H}\right)\, \exp[-y/H]\, ,  
\end{align}
\begin{align} \label{e:BfieldPSy}
    B_{y}(x,y)= & - \frac{2\sigma B_\text{PS} (x - x_\text{PS})^2}{((x - x_\text{PS})^2 + (y - y_\text{PS})^2)^2} + \\ \nonumber
    & \frac{\sigma B_\text{PS}}{(x - x_\text{PS})^2 + (y - y_\text{PS})^2} +  \\ \nonumber
    & B_0\ \cos\left(\frac{x-x_\text{CH}}{H}\right)\,\exp[-y/H] \, , 
\end{align}

\noindent
where $B_\mathrm{PS}=-0.7$ G is the magnetic field strength due to a single line dipole ($\sigma =3\times10^{19}$ is a dimensionless scaling factor) located at $(x, y) = (x_\mathrm{PS}, y_\mathrm{PS})$, where $x_\mathrm{PS}=120\,\mathrm{Mm}$ and $y_\mathrm{PS}=-10\,\mathrm{Mm}$, $B_0=1$ G is the background field strength at $(x, y) = (x_\mathrm{CH}, 0)$, where $x_\mathrm{CH}=200\,\mathrm{Mm}$, and $H=400\,$Mm is the height decay factor. 
The rest of the simulation parameters are set as in \citetalias{Sahade2022} except for the current densities, being here: $j_0=-700~\textrm{statA cm}^{-2},\,j_1=516~\textrm{statA cm}^{-2}$.
\begin{figure}    
   \centerline{\includegraphics[width=0.49\textwidth,clip=]{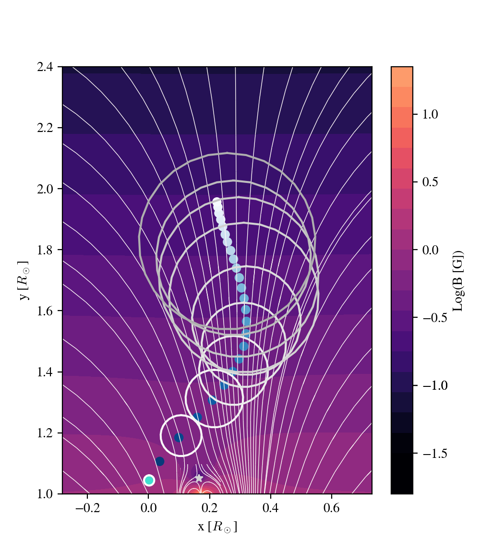}
              }
 \caption{Background magnetic field magnitude and lines of the simulated case. The turquoise dot indicates the initial position of the FR, the gray star indicates the observed PS null point position. The MFR cross section and center represented by white/gray-scaled circles and blue-scaled dots, respectively.}\label{fig:B-simu}
\end{figure}

Figure~\ref{fig:B-simu} shows the background magnetic field resulting from \eqref{e:BfieldPSx}-\eqref{e:BfieldPSy}, where the MFR is added. The turquoise dot and the gray star indicate the prominence initial position and PS2 null point position from the observational data, respectively. The null point position, the width of the PS, and the magnetic field strength are well reproduced (compare to Figure~\ref{fig:B-triang}) by the parameter selection. We note that above $y=2\,R_\sun$ and beyond $x=0.4\,R_\sun$, the field lines behave differently than in Figure~\ref{fig:B-triang}. This is expected since the model assumes a simpler configuration than the actual solar magnetic field. However, it is not necessary to modify the magnetic field configuration to fit those farther regions since our intention is to understand the initial behavior of the MFR. The blue-scaled dots and gray circles show the evolution of the MFR center and cross section, respectively. While the trajectory and the MFR size are comparable, the simulation evolves faster than the observed case. 

\begin{figure}    
   \centerline{\includegraphics[width=0.49\textwidth,clip=]{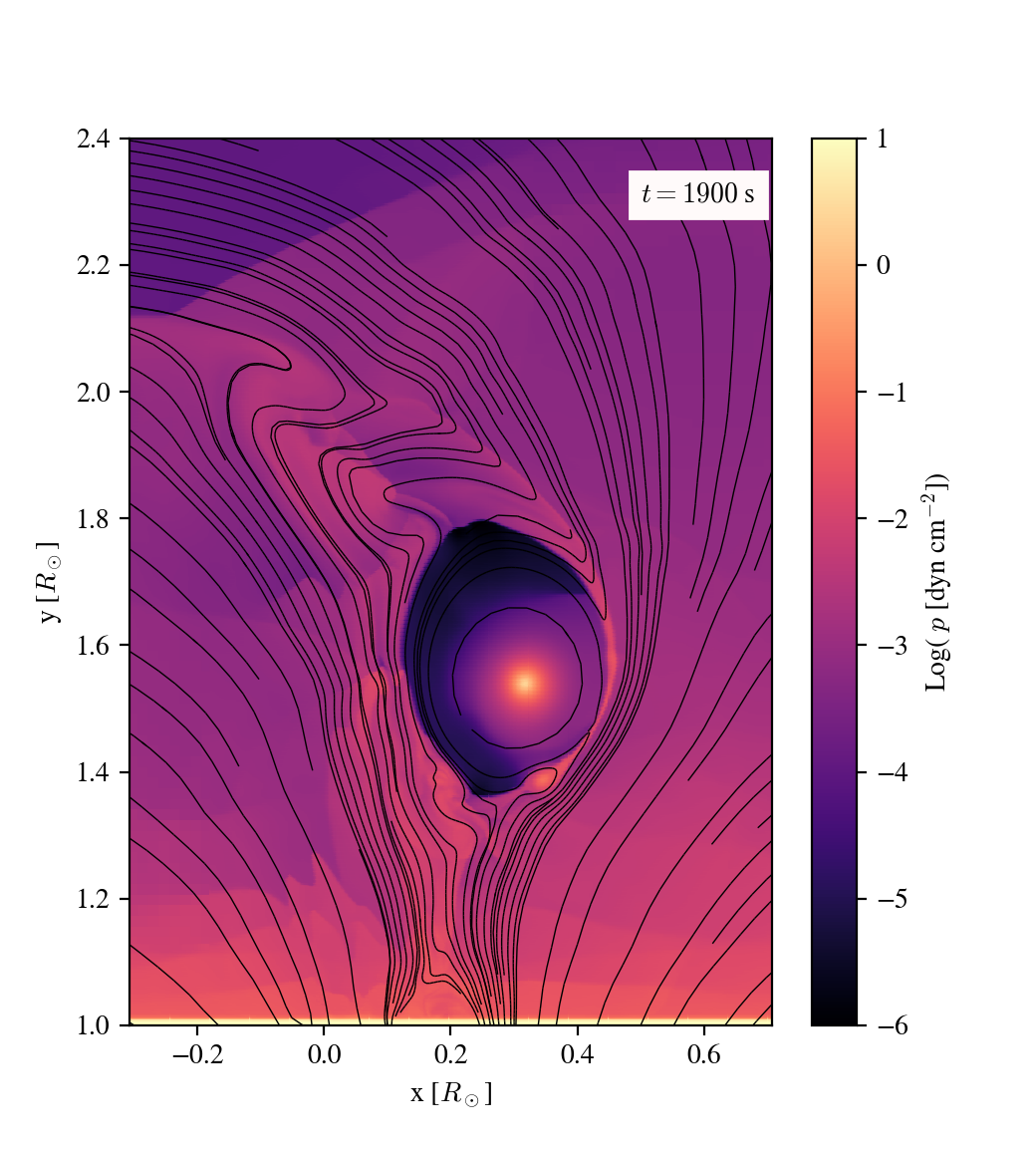}
              }
 \caption{ Gas pressure distribution and magnetic field lines at $t=1900\,$s from the numerical simulation. The field line density is higher in the PS and MFR region to better capture the evolution of these structures ($x\sim[0.1-0.3]\, R_\sun$). An animated version of this figure, showing the magnetic field lines and pressure evolution, is available in the HTML version. The animation shows the MFR evolution during the first $3000\,$s of the simulation, during the first $1900\,$s the MFR evolves towards the right, pushing and bending the CH magnetic field lines, and later the MFR continues its rise moving back to the left. The gas pressure shows the evolution of plasma features such as the MFR, the shock front, and the magnetic islands around the MFR.}  \label{fig:Simu}
\end{figure}

Figure~\ref{fig:Simu} shows the gas pressure distribution and magnetic field lines of the simulation at $t=1900\,$s. We note the MFR (delimited by the low pressure cavity) is displaced to the right and expanded compared to its initial position and volume (see the full evolution in the animated version). During its ascent the MFR develops a super-Alfv\'enic shock ahead of the cavity. The shock interacts with the CH bending it field lines. In the MFR separatrix there is a flux cancellation region (bottom-right diagonal of the MFR) that leads into reconnection and magnetic islands formation. In the animated version, we can follow a magnetic island, located at $(x,y
)\sim (0.4\,R_\sun,1.4\,R_\sun)$ at $t=1500\,$s, as it rotates along the MFR edge almost $90^\circ$ in $900\,$s, and it is lost within the turbulence afterwards.

\begin{figure}    
   \centerline{\includegraphics[width=0.4\textwidth,clip=]{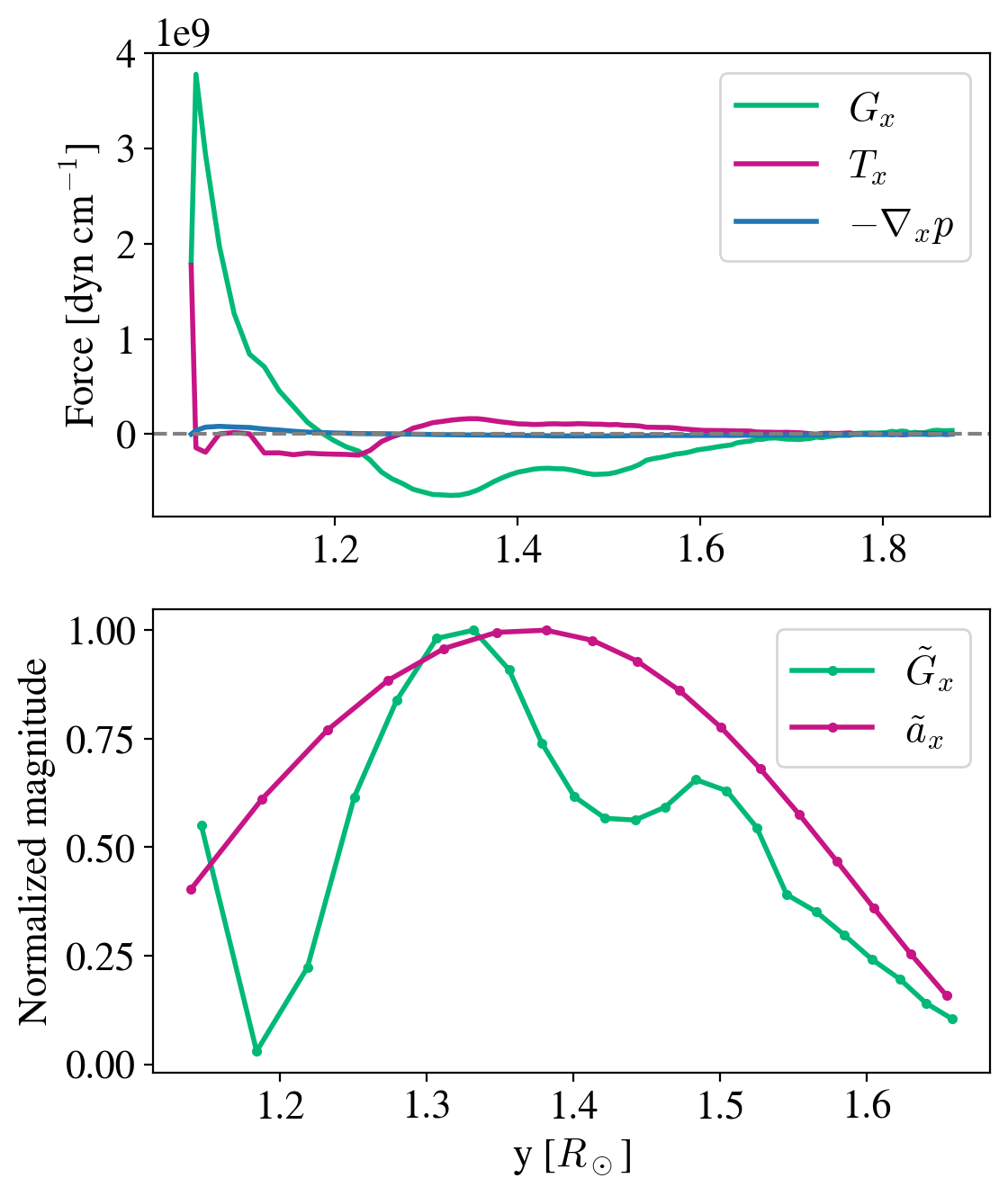}
              }
 \caption{Upper panel: Forces per length unit in $x$-direction exerted on the MFR. Magnetic pressure gradient ($G_x$), magnetic tension ($T_x$) and gas pressure gradient ($-\nabla _x p  $) for the first 4000 s of the simulation. Lower panel: Normalized magnetic gradient pressure ($\tilde{G}_x$) and normalized fit acceleration ($\tilde{a}_x$) in $x$-direction for the simulated MFR during the interaction with the open field lines.}\label{fig:forces}
\end{figure}

The simulated scenario reproduces the behavior observed in the `Cartwheel CME.' The MFR travels toward the null point location and it continues the lateral motion pushing the CH field lines. The CH lines are initially bent by the shock and the MFR, but they eventually stop the rightward motion of the MFR, push back, and guide the MFR back to the original position of the CH lines. 
From the dynamic evolution of the magnetic field, we can calculate the forces exerted on the MFR during its ascent. Figure~\ref{fig:forces} shows the evolution of the MFR forces per length unit (since the simulation is 2.5D) in $x$-direction in the upper panel. The magnetic forces have a larger contribution than the gas pressure, and both magnetic pressure gradient and tension accelerate the MFR to the right (toward the null point) in the initial phase of the evolution. The magnetic forces are driving the MFR toward the null point position. When the MFR begins to interact with the open magnetic field lines of the CH, the magnetic pressure gradient becomes negative, stopping the rightward motion. The maximum MFR displacement occurs at $y \sim 1.6\, R_\sun$ ( and $x \sim 0.3\, R_\sun$ as observed). After that, the negative magnetic pressure gradient increases linearly to zero, as the magnetic tension decreases to zero. Thereafter, the magnetic pressure exerted by the open magnetic field lines restores the MFR to the force-free direction. To compare with Figure~\ref{fig:ax-gradp}, lower panel of Figure~\ref{fig:forces} shows a zoom-in height of the normalized magnetic gradient pressure ($\tilde{G}_x$) and normalized fit acceleration ($\tilde{a}_x$) in $x$-direction during the interaction with the open field lines. For the simulation, in agreement with the observational analysis, the acceleration is directed to the $-x$-direction and gradually reduces to zero. Also, the trend of the magnetic gradient is similar to the acceleration, indicating its contribution in the MFR deceleration during the interaction between the MFR and the CH.

\section{Discussion and Conclusions}\label{sec:conclusion}

The `Cartwheel CME' is a well-studied event with a dynamic behavior that appears to run contrary to the current understanding of the interaction between MFRs and ambient magnetic structures \citep{Capannolo2017}. To investigate whether the CME's behavior was indeed unusual, we first focus on obtaining a more precise reconstruction of the event than previous attempts \citep{Landi2010ApJ,gui2011,Patsourakos2011,Thompson2012}. To achieve this, we use data from three different viewpoints (SOHO, STA and STB) and two different techniques to reconstruct the evolution of different components of the magnetic system within $4\,R_\sun$. Furthermore, we reconstruct the CME using the non-radial GCS model to better capture the non-radial motion of this event. Our reconstructions are consistent with the assumption that the prominence material is located at the bottom of the MFR. The measurements indicate that the prominence undergoes a larger deflection than the MFR center, due possibly to the higher momentum of the heavier prominence material.

Although the deflection produced changes in both latitude and longitude for the prominence apex and MFR, the entire evolution of them can be projected onto a 2D-plane, which simplifies the analysis of the magnetic field configuration in which the MFR moved (see Figure~\ref{fig:3d-view} and \ref{fig:B-triang}). As established before \citep{Savage2010,Capannolo2017}, the MFR was interacting with the open magnetic field lines of a CH, traveling toward them, and against the magnetic gradients. Our thorough reconstruction of the initial rising phase and magnetic field allows the identification of a pseudostreamer (PS2) null point located between the initial position of the MFR and the CH. Considering the action of null points in trajectory \citep[e.g.,][]{Panasenco2013SoPh,Wang2020JGRA,Sahade2021,Sahade2022}, we presume that this null point was responsible for attracting the MFR toward the CH, which then stopped the MFR deviation and guided it parallel to its magnetic field lines. Then, the peculiar behavior of the 2008 April 09 CME can be explained not only by assuming an asymmetric magnetic reconnection \citep{Capannolo2017}, but as a response to the interaction with the magnetic environment near the source region. 

We analyze the alignment between the trajectory and the magnetic pressure gradient (see Figure~\ref{fig:angle}) and observe different phases in the evolution. Initially, as expected, the angle is small as the MFR traveled toward the null point, then it increases abruptly as the MFR crossed the null point location. Once inside the CH, the MFR trajectory smoothly aligned with both the magnetic gradient and magnetic fields. Finally, we see that the misalignment between the angles increase, presumably because the CH is reacting to the MFR displacement and pushing it back toward a more radial path. In conclusion, our analysis indicates that the MFR tried to align with the magnetic pressure gradient. However, it should be noted that null points can lead to stronger deflections producing misalignment and also, at later evolution, the angles calculated from the static magnetic field extrapolation may not reflect the responses of the magnetic structure. We estimate the dynamic magnetic pressure gradient exerted by the CH on the MFR (Figure~\ref{fig:ax-gradp}) and find that it correlates with the non-radial acceleration of the MFR. Consequently, we conclude that the magnetic pressure gradient is at least one of the restorative forces producing the reversal deflection. 

We perform ideal MHD-simulations to model the dynamics of the 2008 April 09 event, adapting the magnetic scenario explored in \citetalias{Sahade2022}. The simulation considers a MFR interacting with a PS structure similar to the observed one (see Figures~\ref{fig:B-simu} and \ref{fig:Simu}), other magnetic structures as the nearby AR are excluded from the modeling since they are not contained in the POE. The simulated MFR presents the same double-deflection behavior as the `Cartwheel CME,' validating the relevance of the null point and magnetic configuration. We calculate the forces acting over the MFR by considering the dynamic evolution of the environment. Initially, the magnetic tension and the magnetic pressure gradient are responsible for deviating the MFR in a non-radial direction and toward the null point. Then, the magnetic pressure gradient is the restoring force that stops the MFR deflection and pushes it back toward a direction more aligned with the original CH lines, agreeing with the data analysis. 

In summary, we find, observationally and numerically, that the behavior of the `Cartwheel CME' can be explained once the trajectory and magnetic environment are well described. The evolution can be divided into three phases. The first one is driven by the presence of the PS2 null point (deflection to the south until 10:09 UT), the second one consists of the response to the CH (reversal deflection), and the third one concerns the MFR outward propagation parallel to the magnetic field lines following the least resistance path (near radial trajectory after 11:05 UT).

The most important conclusions  drawn from this work are:
\begin{itemize}
    \item The dynamic behavior of the CME was not unusual but rather as expected. The CME escapes through the nearest null point as expected on physical considerations. The apparent 'rolling' behavior and sharp direction change were due to the topological configuration in the vicinity of the eruption. 
    \item Multi-viewpoint observations of the low coronal evolution of an eruption are key for understanding the topological environment around the erupting MFR. They can provide essential information to understand unexpected behaviors.
    \item Null points play a key role in the early evolution of erupting MFRs. Identifying their presence and, more generally, determining the ambient magnetic topology, will greatly improve our understanding of the early development and trajectory of eruptions.
    \item 2.5 MHD numerical simulations provide a useful tool to study different scenarios in which a MFR can evolve. They are computationally less expensive than, for example, data-driven models and allow us to test the interpretations that can not be easily verified with data. 

\end{itemize}
Recent developments in instrumentation and observations promise great opportunities for further understanding the early evolution of CMEs. EUV and white observations from Solar Orbiter provide a `3rd' eye to the observations from STEREO, and Earth-based assets (e.g SDO, GOES/SUVI and SWFO-L1 in 2025+) from widely variable viewpoints. The future addition of off Sun-Earth line magnetograms (via ESA's Vigil mission, currently in development) will further enhance the reliability of magnetic field extrapolations and consequently of topological maps of the solar corona.  

\begin{acknowledgements}
    AS is doctoral fellow of CONICET. AS, AV and LAB are supported by NASA grant 80NSSC19K0069. MC is member of the Carrera del Investigador Cient\'ifico (CONICET). AS and MC acknowledge support from SECYT-UNC grant number PC No. 33620180101147CB, and support from PIP under grant number No. 11220200103150CO. We also thank the Centro de C\'omputo de Alto Desempe\~no (UNC), where the simulations were carried out. The work presented here was carried out at JHU/APL and GMU as part of a research internship. AS thanks JHU/APL and GMU for their hospitality during her visit. 
 \end{acknowledgements}

\appendix

\section{Prominence and CME 3D reconstruction}\label{app:tabla}
Table~\ref{tbl-coord} displays the 3D-coordinates for each timestamp with the different techniques, instruments and features of the erupting structure.
\begin{deluxetable*}{c | c c c | c c c | c c c}
\tablecaption{Spherical coordinates of the prominence apex, CME center, and CME apex.\label{tbl-coord}}

\tablehead{
\colhead{Time [UT]}  & \multicolumn{3}{c}{Prominence apex}& \multicolumn{3}{c}{CME center} & \multicolumn{3}{c}{CME apex}\\
\nocolhead{}& \colhead{R [$R_\sun$]}& \colhead{$\theta$ [$^\circ$]}& \colhead{$\phi$ [$^\circ$]}& \colhead{R [$R_\sun$]}& \colhead{$\theta$ [$^\circ$]}& \colhead{$\phi$ [$^\circ$]}& \colhead{R [$R_\sun$]}& \colhead{$\theta$ [$^\circ$]}& \colhead{$\phi$ [$^\circ$]}
}

\startdata
  08:15& $1.03$*& $-16.2$& $195.4$& \multicolumn{6}{c}{-} \\
  08:25& $1.03$*& $-16.4$& $195.4$& \multicolumn{6}{c}{-} \\
  08:45& $1.03$*& $-16.6$& $196.2$& \multicolumn{6}{c}{-} \\
  09:15& $1.08$*& $-19.7$& $194.7$& \multicolumn{6}{c}{-} \\
  09:25& $1.11$*& $-21.6$& $195.1$& $1.24$& $-22.1$& $188.4$& $1.29$& $-23.0$& $187.5$ \\
  09:35& $1.13$*& $-22.3$& $191.8$& $1.24$& $-22.2$& $188.3$& $1.30$& $-23.3$& $187.2$ \\
  09:45& $1.18$*& $-22.9$& $186.5$& $1.27$& $-22.8$& $187.6$& $1.36$& $-24.1$& $186.2$ \\
  09:48& $1.18$ & $-22.9$& $186.5$& $1.29$& $-23.0$& $187.4$& $1.38$& $-24.3$& $186.0$ \\
  09:51& $1.18$& $-24.6$& $186.0$& \multicolumn{6}{c}{-} \\
  09:53& $1.19$& $-25.5$& $185.9$& \multicolumn{6}{c}{-} \\
  09:56& $1.24$& $-26.1$& $184.4$& \multicolumn{6}{c}{-} \\
  09:58& $1.21$& $-26.3$& $188.4$& \multicolumn{6}{c}{-} \\
  10:01& $1.31$& $-27.8$& $181.8$& $1.29$& $-23.2$& $187.3$& $1.40$& $-24.8$& $185.5$  \\
  10:03& $1.31$& $-28.9$& $186.4$& \multicolumn{6}{c}{-} \\
  10:06& $1.32$& $-28.2$& $184.0$& $1.38$& $-23.3$& $187.1$& $1.54$& $-24.9$& $185.3$  \\
  10:08& $1.39$& $-28.8$& $182.2$& \multicolumn{6}{c}{-} \\
  10:09& $1.43$& $-28.5$& $183.9$& \multicolumn{6}{c}{-} \\
  10:11& $1.47$& $-28.3$& $181.8$& \multicolumn{6}{c}{-} \\
  10:13& $1.52$& $-28.3$& $182.0$& $1.56$& $-25.1$& $185.1$& $1.78$& $-26.7$& $183.3$  \\
  10:16& $1.54$& $-29.3$& $186.1$& \multicolumn{6}{c}{-} \\
  10:18& $1.60$& $-28.9$& $184.4$& \multicolumn{6}{c}{-} \\
  10:09& $1.64$& $-29.6$& $182.6$& \multicolumn{6}{c}{-} \\
  10:21& $1.71$& $-29.4$& $183.3$& \multicolumn{6}{c}{-} \\
  10:23& $1.75$& $-29.0$& $181.9$& \multicolumn{6}{c}{-} \\
  10:26& $1.86$& $-27.3$& $183.1$& $1.81$& $-25.4$& $184.8$& $2.05$& $-26.6$& $183.4$  \\
  10:40&\multicolumn{3}{c}{-}    & $2.08$& $-22.9$& $187.6$& $2.36$& $-23.5$& $186.9$ \\
  \hline
  10:45& $2.25$& $-26.3$& $192.4$& $2.21$& $-21.5$& $189.0$& $2.56$& $-22.0$& $188.5$  \\
  10:55& $2.47$& $-22.8$& $191.3$& $2.45$& $-21.0$& $189.5$& $2.91$& $-21.5$& $189.1$  \\
  11:05& $2.80$& $-25.2$& $194.5$& $2.76$& $-19.6$& $190.9$& $3.30$& $-21.5$& $190.7$ \\
  11:15& $3.08$& $-26.2$& $195.7$& $2.81$& $-19.6$& $190.9$& $3.36$& $-19.9$& $190.7$ \\
  11:20& \multicolumn{3}{c}{-}   & $3.09$& $-19.8$& $190.8$& $3.70$& $-20.0$& $190.6$ \\
  11:25& \multicolumn{3}{c}{-}   & $3.24$& $-18.9$& $191.7$& $3.87$& $-19.0$& $191.5$ \\
  11:45& \multicolumn{3}{c}{-}   & $4.10$& $-19.1$& $191.5$& $4.91$& $-19.2$& $191.4$  
\enddata
\tablecomments{The coordinates of the prominence apex are determined with the tie-pointing technique, using STA and STB spacecrafts (* use STA and SOHO). The coordinates of the CME (center and apex) are determined from the non-radial GCS reconstruction, using STA, STB and SOHO images. Times until 10:40 UT correspond to EUV images, times from 10:45 correspond to coronagraph images.}
\end{deluxetable*}

\bibliography{biblio}{}

\begin{thebibliography}{}
\expandafter\ifx\csname natexlab\endcsname\relax\def\natexlab#1{#1}\fi
\providecommand{\url}[1]{\href{#1}{#1}}
\providecommand{\dodoi}[1]{doi:~\href{http://doi.org/#1}{\nolinkurl{#1}}}
\providecommand{\doeprint}[1]{\href{http://ascl.net/#1}{\nolinkurl{http://ascl.net/#1}}}
\providecommand{\doarXiv}[1]{\href{https://arxiv.org/abs/#1}{\nolinkurl{https://arxiv.org/abs/#1}}}

\bibitem[{{Brueckner} {et~al.}(1995){Brueckner}, {Howard}, {Koomen},
  {Korendyke}, {Michels}, {Moses}, {Socker}, {Dere}, {Lamy}, {Llebaria},
  {Bout}, {Schwenn}, {Simnett}, {Bedford}, \& {Eyles}}]{1995SoPhLASCO}
{Brueckner}, G.~E., {Howard}, R.~A., {Koomen}, M.~J., {et~al.} 1995, \solphys,
  162, 357, \dodoi{10.1007/BF00733434}

\bibitem[{{Capannolo} {et~al.}(2017){Capannolo}, {Opher}, {Kay}, \&
  {Landi}}]{Capannolo2017}
{Capannolo}, L., {Opher}, M., {Kay}, C., \& {Landi}, E. 2017, \apj, 839, 37,
  \dodoi{10.3847/1538-4357/aa6a16}

\bibitem[{{C{\'e}cere} {et~al.}(2020){C{\'e}cere}, {Sieyra}, {Cremades},
  {Mierla}, {Sahade}, {Stenborg}, {Costa}, {West}, \& {D'Huys}}]{cecere2020}
{C{\'e}cere}, M., {Sieyra}, M.~V., {Cremades}, H., {et~al.} 2020, Advances in
  Space Research, 65, 1654, \dodoi{10.1016/j.asr.2019.08.043}

\bibitem[{{Cremades} \& {Bothmer}(2004)}]{cremades2004}
{Cremades}, H., \& {Bothmer}, V. 2004, \aap, 422, 307,
  \dodoi{10.1051/0004-6361:20035776}

\bibitem[{{Cremades} {et~al.}(2006){Cremades}, {Bothmer}, \&
  {Tripathi}}]{Cremades2006}
{Cremades}, H., {Bothmer}, V., \& {Tripathi}, D. 2006, Advances in Space
  Research, 38, 461, \dodoi{10.1016/j.asr.2005.01.095}

\bibitem[{{Delaboudini{\`e}re} {et~al.}(1995){Delaboudini{\`e}re}, {Artzner},
  {Brunaud}, {Gabriel}, {Hochedez}, {Millier}, {Song}, {Au}, {Dere}, {Howard},
  {Kreplin}, {Michels}, {Moses}, {Defise}, {Jamar}, {Rochus}, {Chauvineau},
  {Marioge}, {Catura}, {Lemen}, {Shing}, {Stern}, {Gurman}, {Neupert},
  {Maucherat}, {Clette}, {Cugnon}, \& {Van Dessel}}]{1995SoPhEIT}
{Delaboudini{\`e}re}, J.~P., {Artzner}, G.~E., {Brunaud}, J., {et~al.} 1995,
  \solphys, 162, 291, \dodoi{10.1007/BF00733432}

\bibitem[{{Domingo} {et~al.}(1995){Domingo}, {Fleck}, \&
  {Poland}}]{1995SoPh..162....1D}
{Domingo}, V., {Fleck}, B., \& {Poland}, A.~I. 1995, \solphys, 162, 1,
  \dodoi{10.1007/BF00733425}

\bibitem[{{Filippov}(2019)}]{Filippov2019}
{Filippov}, B.~P. 2019, Physics Uspekhi, 62, 847,
  \dodoi{10.3367/UFNe.2018.10.038467}

\bibitem[{{Gopalswamy} {et~al.}(2009){Gopalswamy}, {M{\"a}kel{\"a}}, {Xie},
  {Akiyama}, \& {Yashiro}}]{Gopalswamy2009}
{Gopalswamy}, N., {M{\"a}kel{\"a}}, P., {Xie}, H., {Akiyama}, S., \& {Yashiro},
  S. 2009, Journal of Geophysical Research (Space Physics), 114, A00A22,
  \dodoi{10.1029/2008JA013686}

\bibitem[{{Green} {et~al.}(2018){Green}, {T{\"o}r{\"o}k}, {Vr{\v{s}}nak},
  {Manchester}, \& {Veronig}}]{Green2018}
{Green}, L.~M., {T{\"o}r{\"o}k}, T., {Vr{\v{s}}nak}, B., {Manchester}, W., \&
  {Veronig}, A. 2018, \ssr, 214, 46, \dodoi{10.1007/s11214-017-0462-5}

\bibitem[{{Gui} {et~al.}(2011){Gui}, {Shen}, {Wang}, {Ye}, {Liu}, {Wang}, \&
  {Zhao}}]{gui2011}
{Gui}, B., {Shen}, C., {Wang}, Y., {et~al.} 2011, \solphys, 271, 111,
  \dodoi{10.1007/s11207-011-9791-9}

\bibitem[{{Howard} {et~al.}(2008){Howard}, {Moses}, {Vourlidas}, {Newmark},
  {Socker}, {Plunkett}, {Korendyke}, {Cook}, {Hurley}, {Davila}, \& {et
  al.}}]{Howard2008}
{Howard}, R.~A., {Moses}, J.~D., {Vourlidas}, A., {et~al.} 2008, \ssr, 136, 67,
  \dodoi{10.1007/s11214-008-9341-4}

\bibitem[{{Inhester}(2006)}]{Inhester2006}
{Inhester}, B. 2006, arXiv Astrophysics e-prints

\bibitem[{{Isavnin}(2016)}]{Isavnin2016}
{Isavnin}, A. 2016, \apj, 833, 267, \dodoi{10.3847/1538-4357/833/2/267}

\bibitem[{{Jiang} {et~al.}(2018){Jiang}, {Feng}, \& {Hu}}]{Jiang2018}
{Jiang}, C., {Feng}, X., \& {Hu}, Q. 2018, \apj, 866, 96,
  \dodoi{10.3847/1538-4357/aadd08}

\bibitem[{{Kaiser} {et~al.}(2008){Kaiser}, {Kucera}, {Davila}, {St.~Cyr},
  {Guhathakurta}, \& {Christian}}]{kaiser2008}
{Kaiser}, M.~L., {Kucera}, T.~A., {Davila}, J.~M., {et~al.} 2008, Space Sci.
  Rev., 136, 5, \dodoi{10.1007/s11214-007-9277-0}

\bibitem[{{Karna} {et~al.}(2021){Karna}, {Savcheva}, {Gibson}, {Tassev},
  {Reeves}, {DeLuca}, \& {Dalmasse}}]{Karna2021}
{Karna}, N., {Savcheva}, A., {Gibson}, S., {et~al.} 2021, \apj, 913, 47,
  \dodoi{10.3847/1538-4357/abf2b8}

\bibitem[{{Kay} {et~al.}(2013){Kay}, {Opher}, \& {Evans}}]{Kay2013}
{Kay}, C., {Opher}, M., \& {Evans}, R.~M. 2013, \apj, 775, 5,
  \dodoi{10.1088/0004-637X/775/1/5}

\bibitem[{{Kay} {et~al.}(2015){Kay}, {Opher}, \& {Evans}}]{kay2015}
---. 2015, \apj, 805, 168, \dodoi{10.1088/0004-637X/805/2/168}

\bibitem[{{Kliem} {et~al.}(2012){Kliem}, {T{\"o}r{\"o}k}, \&
  {Thompson}}]{Kliem2012}
{Kliem}, B., {T{\"o}r{\"o}k}, T., \& {Thompson}, W.~T. 2012, \solphys, 281,
  137, \dodoi{10.1007/s11207-012-9990-z}

\bibitem[{{Kwon} {et~al.}(2014){Kwon}, {Zhang}, \& {Olmedo}}]{Kwon2014}
{Kwon}, R.-Y., {Zhang}, J., \& {Olmedo}, O. 2014, \apj, 794, 148,
  \dodoi{10.1088/0004-637X/794/2/148}

\bibitem[{{Landi} {et~al.}(2010){Landi}, {Raymond}, {Miralles}, \&
  {Hara}}]{Landi2010ApJ}
{Landi}, E., {Raymond}, J.~C., {Miralles}, M.~P., \& {Hara}, H. 2010, \apj,
  711, 75, \dodoi{10.1088/0004-637X/711/1/75}

\bibitem[{{Liewer} {et~al.}(2015){Liewer}, {Panasenco}, {Vourlidas}, \&
  {Colaninno}}]{Liewer2015}
{Liewer}, P., {Panasenco}, O., {Vourlidas}, A., \& {Colaninno}, R. 2015,
  \solphys, 290, 3343, \dodoi{10.1007/s11207-015-0794-9}

\bibitem[{{MacQueen} {et~al.}(1986){MacQueen}, {Hundhausen}, \&
  {Conover}}]{macqueen1986}
{MacQueen}, R.~M., {Hundhausen}, A.~J., \& {Conover}, C.~W. 1986, \jgr, 91, 31,
  \dodoi{10.1029/JA091iA01p00031}

\bibitem[{{Maloney} {et~al.}(2009){Maloney}, {Gallagher}, \&
  {McAteer}}]{maloney2009}
{Maloney}, S.~A., {Gallagher}, P.~T., \& {McAteer}, R.~T.~J. 2009, \solphys,
  256, 149, \dodoi{10.1007/s11207-009-9364-3}

\bibitem[{{Mierla} {et~al.}(2008){Mierla}, {Davila}, {Thompson}, {Inhester},
  {Srivastava}, {Kramar}, {St.~Cyr}, {Stenborg}, \& {Howard}}]{mierla2008}
{Mierla}, M., {Davila}, J., {Thompson}, W., {et~al.} 2008, \solphys, 252, 385,
  \dodoi{10.1007/s11207-008-9267-8}

\bibitem[{{M{\"o}stl} {et~al.}(2015){M{\"o}stl}, {Rollett}, {Frahm}, {Liu},
  {Long}, {Colaninno}, {Reiss}, {Temmer}, {Farrugia}, {Posner}, {Dumbovi{\'c}},
  {Janvier}, {D{\'e}moulin}, {Boakes}, {Devos}, {Kraaikamp}, {Mays}, \& {Vr{\v
  s}nak}}]{Mostl2015}
{M{\"o}stl}, C., {Rollett}, T., {Frahm}, R.~A., {et~al.} 2015, Nature
  Communications, 6, 7135, \dodoi{10.1038/ncomms8135}

\bibitem[{{Panasenco} {et~al.}(2013){Panasenco}, {Martin}, {Velli}, \&
  {Vourlidas}}]{Panasenco2013SoPh}
{Panasenco}, O., {Martin}, S.~F., {Velli}, M., \& {Vourlidas}, A. 2013,
  \solphys, 287, 391, \dodoi{10.1007/s11207-012-0194-3}

\bibitem[{{Patsourakos} \& {Vourlidas}(2011)}]{Patsourakos2011}
{Patsourakos}, S., \& {Vourlidas}, A. 2011, \aap, 525, A27,
  \dodoi{10.1051/0004-6361/201015048}

\bibitem[{{Sahade} {et~al.}(2021){Sahade}, {C{\'e}cere}, {Costa}, \&
  {Cremades}}]{Sahade2021}
{Sahade}, A., {C{\'e}cere}, M., {Costa}, A., \& {Cremades}, H. 2021, \aap, 652,
  A111, \dodoi{10.1051/0004-6361/202141085}

\bibitem[{{Sahade} {et~al.}(2020){Sahade}, {C{\'e}cere}, \&
  {Krause}}]{Sahade2020}
{Sahade}, A., {C{\'e}cere}, M., \& {Krause}, G. 2020, \apj, 896, 53,
  \dodoi{10.3847/1538-4357/ab8f25}

\bibitem[{{Sahade} {et~al.}(2022){Sahade}, {C{\'e}cere}, {Sieyra}, {Krause},
  {Cremades}, \& {Costa}}]{Sahade2022}
{Sahade}, A., {C{\'e}cere}, M., {Sieyra}, M.~V., {et~al.} 2022, \aap, 662,
  A113, \dodoi{10.1051/0004-6361/202243618}

\bibitem[{{Savage} {et~al.}(2010){Savage}, {McKenzie}, {Reeves}, {Forbes}, \&
  {Longcope}}]{Savage2010}
{Savage}, S.~L., {McKenzie}, D.~E., {Reeves}, K.~K., {Forbes}, T.~G., \&
  {Longcope}, D.~W. 2010, \apj, 722, 329, \dodoi{10.1088/0004-637X/722/1/329}

\bibitem[{{Scherrer} {et~al.}(1995){Scherrer}, {Bogart}, {Bush}, {Hoeksema},
  {Kosovichev}, {Schou}, {Rosenberg}, {Springer}, {Tarbell}, {Title},
  {Wolfson}, {Zayer}, \& {MDI Engineering Team}}]{1995SoPh..162..129S}
{Scherrer}, P.~H., {Bogart}, R.~S., {Bush}, R.~I., {et~al.} 1995, \solphys,
  162, 129, \dodoi{10.1007/BF00733429}

\bibitem[{{Schrijver} \& {De Rosa}(2003)}]{2003SoPh..212..165S}
{Schrijver}, C.~J., \& {De Rosa}, M.~L. 2003, \solphys, 212, 165,
  \dodoi{10.1023/A:1022908504100}

\bibitem[{{Sieyra} {et~al.}(2020){Sieyra}, {C{\'e}cere}, {Cremades},
  {Iglesias}, {Sahade}, {Mierla}, {Stenborg}, {Costa}, {West}, \&
  {D'Huys}}]{Sieyra2020}
{Sieyra}, M.~V., {C{\'e}cere}, M., {Cremades}, H., {et~al.} 2020, \solphys,
  295, 126, \dodoi{10.1007/s11207-020-01694-0}

\bibitem[{{Temmer} {et~al.}(2009){Temmer}, {Preiss}, \& {Veronig}}]{temmer2009}
{Temmer}, M., {Preiss}, S., \& {Veronig}, A.~M. 2009, \solphys, 256, 183,
  \dodoi{10.1007/s11207-009-9336-7}

\bibitem[{{Thernisien} {et~al.}(2009){Thernisien}, {Vourlidas}, \&
  {Howard}}]{Thernisien2009}
{Thernisien}, A., {Vourlidas}, A., \& {Howard}, R.~A. 2009, \solphys, 256, 111,
  \dodoi{10.1007/s11207-009-9346-5}

\bibitem[{{Thompson} {et~al.}(2012){Thompson}, {Kliem}, \&
  {T{\"o}r{\"o}k}}]{Thompson2012}
{Thompson}, W.~T., {Kliem}, B., \& {T{\"o}r{\"o}k}, T. 2012, \solphys, 276,
  241, \dodoi{10.1007/s11207-011-9868-5}

\bibitem[{{van Driel-Gesztelyi} \& {Green}(2015)}]{vanDriel2015}
{van Driel-Gesztelyi}, L., \& {Green}, L.~M. 2015, Living Reviews in Solar
  Physics, 12, 1, \dodoi{10.1007/lrsp-2015-1}

\bibitem[{Vourlidas {et~al.}(2013)Vourlidas, Lynch, Howard, \&
  Li}]{Vourlidas_etal_2013}
Vourlidas, A., Lynch, B.~J., Howard, R.~A., \& Li, Y. 2013, Solar Physics, 284,
  179–201, \dodoi{10.1007/s11207-012-0084-8}

\bibitem[{{Wang} {et~al.}(2020){Wang}, {Hoeksema}, \& {Liu}}]{Wang2020JGRA}
{Wang}, J., {Hoeksema}, J.~T., \& {Liu}, S. 2020, Journal of Geophysical
  Research (Space Physics), 125, e27530, \dodoi{10.1029/2019JA027530}

\bibitem[{{Wang} {et~al.}(2015){Wang}, {Liu}, {Dai}, {Yang}, {Huang}, \&
  {Hu}}]{Wang2015}
{Wang}, R., {Liu}, Y.~D., {Dai}, X., {et~al.} 2015, \apj, 814, 80,
  \dodoi{10.1088/0004-637X/814/1/80}

\bibitem[{{Yang} {et~al.}(2018){Yang}, {Dai}, {Chen}, {Li}, \&
  {Jiang}}]{Yang2018}
{Yang}, J., {Dai}, J., {Chen}, H., {Li}, H., \& {Jiang}, Y. 2018, \apj, 862,
  86, \dodoi{10.3847/1538-4357/aaccfd}

\bibitem[{{Zhang} {et~al.}(2001){Zhang}, {Dere}, {Howard}, {Kundu}, \&
  {White}}]{Zhang2001}
{Zhang}, J., {Dere}, K.~P., {Howard}, R.~A., {Kundu}, M.~R., \& {White}, S.~M.
  2001, \apj, 559, 452, \dodoi{10.1086/322405}

\bibitem[{{Zhang}(2021)}]{Zhang2021A&A}
{Zhang}, Q.~M. 2021, \aap, 653, L2, \dodoi{10.1051/0004-6361/202141982}

\bibitem[{{Zuccarello} {et~al.}(2012){Zuccarello}, {Bemporad}, {Jacobs},
  {Mierla}, {Poedts}, \& {Zuccarello}}]{Zuccarello2012}
{Zuccarello}, F.~P., {Bemporad}, A., {Jacobs}, C., {et~al.} 2012, \apj, 744,
  66, \dodoi{10.1088/0004-637X/744/1/66}

\end{thebibliography}
\bibliographystyle{aasjournal}

\end{document}